\documentclass[aps,prl,twocolumn,groupedaddress,superscriptaddress,floatfix]{revtex4-1}
\usepackage{graphicx}
\usepackage{dcolumn}
\usepackage{bm}
\usepackage{natbib}
\usepackage{amsmath}
\usepackage{amssymb}
\usepackage{hyperref}
\usepackage{color}
\newcommand{\eref}[1]{Eq.\,\eqref{#1}}
\newcommand{\mc}[1]{\mathcal{#1}}

\makeatother

\begin{document}
\title{2D Superexchange mediated magnetization dynamics in an optical lattice}

\author{R. C. Brown}
\affiliation{Joint Quantum Institute, National Institute of Standards and Technology, and University of Maryland, Gaithersburg, Maryland 20899, USA}

\author{R. Wyllie}
\altaffiliation{Current address: Quantum Systems Division, Georgia Tech Research Institute, Atlanta, Georgia 30332, USA}
\affiliation{Joint Quantum Institute, National Institute of Standards and Technology, and University of Maryland, Gaithersburg, Maryland 20899, USA}

\author{S. B. Koller}
\affiliation{Joint Quantum Institute, National Institute of Standards and Technology, and University of Maryland, Gaithersburg, Maryland 20899, USA}

\author{E. A. Goldschmidt}
\affiliation{Joint Quantum Institute, National Institute of Standards and Technology, and University of Maryland, Gaithersburg, Maryland 20899, USA}

\author{M. Foss-Feig}
\affiliation{Joint Quantum Institute, National Institute of Standards and Technology, and University of Maryland, Gaithersburg, Maryland 20899, USA}

\author{J. V. Porto}
\affiliation{Joint Quantum Institute, National Institute of Standards and Technology, and University of Maryland, Gaithersburg, Maryland 20899, USA}

\begin{abstract}
The competition of magnetic exchange interactions and tunneling underlies many complex quantum phenomena observed in real materials.  We study non-equilibrium magnetization dynamics in an extended 2D system by loading effective spin-1/2 bosons into a spin-dependent optical lattice, and we use the lattice to separately control the resonance conditions for tunneling and superexchange.  After preparing a non-equilibrium anti-ferromagnetically ordered state, we observe relaxation dynamics governed by two well-separated rates, which scale with the underlying Hamiltonian parameters associated with superexchange and tunneling.  Remarkably, with tunneling off-resonantly suppressed, we are able to observe superexchange dominated dynamics over two orders of magnitude in magnetic coupling strength, despite the presence of vacancies.  In this regime, the measured timescales are in agreement with simple theoretical estimates, but the detailed dynamics of this 2D, strongly correlated, and far-from-equilibrium quantum system remain out of reach of current computational techniques.
\end{abstract}

\date{\today}
\maketitle

The interplay of spin and motion underlies some of the most intriguing and poorly understood behaviors in many-body quantum systems~\cite{Interacting_e_Auerbach1994}. A well known example is the onset of superconductivity in cuprate compounds when mobile holes are introduced into an otherwise insulating 2D quantum magnet~\cite{Acommonthread_RMP_2012}; understanding this behavior is particularly challenging because the dimensionality is low enough to support strong quantum correlations, but high enough to prohibit numerical solution. Ultracold atoms in optical lattices realize tunable, idealized models of such behavior, and can naturally operate in a regime where the quantum motion (tunneling) of particles and magnetic interactions (superexchange) explicitly compete~\cite{JakschPRL1998BHH, SpinModelDuanPRL2003}.

For ultracold atoms in equilibrium, the extremely small energy scale associated with superexchange interactions makes the observation of magnetism challenging, and short-range antiferromagnetic correlations resulting from superexchange have only recently been observed~\cite{Greif_Science2013,Hart_Arxiv_2014}. Out of equilibrium, superexchange-dominated dynamics has been demonstrated in isolated pairs of atoms~\cite{TrotzkyScience2008}, in 1D systems with single atom spin impurities~\cite{Magnon_Boundstates_Fukuhara_Nat_2013Nat}, and recently in the decay of spin-density waves~\cite{HildPRL2014}. However, the perturbative origin of superexchange in these systems requires that it be weak compared to tunneling, and thus the manifestation of superexchange requires extremely low motional entropy. Dipolar gases~\cite{NeqQMCrPRL2013} and ultracold polar molecules~\cite{MoleculeSpinDynamicsYanNature2013} in lattices provide a promising route toward achieving large (non-perturbative) magnetic interactions~\cite{GorshkovPRL2011}, but, technical limitations in these systems currently complicate the simultaneous observation of motional and spin-exchange effects.

 Here, we study the magnetization dynamics of effective spin-1/2 bosons in a 2D optical lattice following a global quench from an initially antiferromagnetically ordered state~\cite{CritSpeedupAFdecayBarmettlerPRL2009}. The dynamics we observe is governed by a bosonic $t$-$J$ model~\cite{tJ_phase_separation_PRL2001,CounterFlowSF_SvistunovPRL}. Utilizing a checkerboard optical lattice, we continuously tune the magnetization dynamics from a tunneling-dominated regime into a regime where superexchange is dominant, even at relatively high motional entropies. This experiment bridges the gap between experiments studying the non-equilibrium behavior of systems with exclusively motional~\cite{CheneauNatureLightcone2012,PrethermalizationGringSci2012,emerg_thermalcorr_Langen_NatPh2013,KinoshitaWeissNature2006QuantumNewtonCradle,StrohmaierPRL2010ElasticDoublonDecay} or spin degrees~\cite{Lightcone_Richerme_Nat_2014,IonDynamicsBlatt2014Nature} of freedom, demonstrating the requisite control to explore the intriguing intermediate territory in which they compete. In addition, the techniques we demonstrate lay the groundwork for adiabatic preparation of low entropy spin states relevant for studies of equilibrium quantum magnetism~\cite{SorensenPRA2010,LubaschPRL2011}.

\begin{figure*}
\centering\includegraphics [width=6.2in,angle=0] {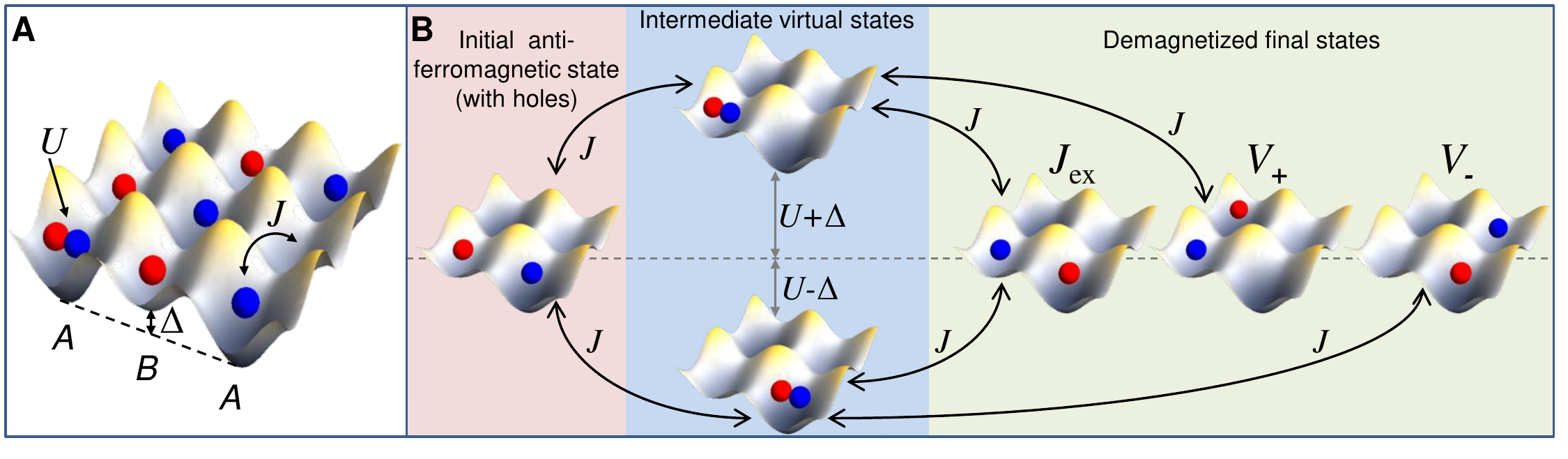}
\caption{ {\bf Tunable exchange processes.} {\bf (A)} Schematic of terms in the underlying Bose-Hubbard Hamiltonian: onsite interaction energy between two atoms $U$, tunneling $J$, and sub lattice offset, $\Delta$.  {\bf (B)} Second order magnetic coupling processes arising from exchange between occupied nearest neighbor sites ($J_{\rm{ex}}$) or hole-mediated exchange associated with hopping of a hole within one sub-lattice ($V_{+}$ and $V_{-}$).  These couplings dominate the magnetization dynamics when tunneling is suppressed by tuning $|\Delta| \gg J$.}
\label{fig:FancyLatticeFig}
\end{figure*}

\begin{figure*}
\centering\includegraphics [width=6.2in,angle=0] {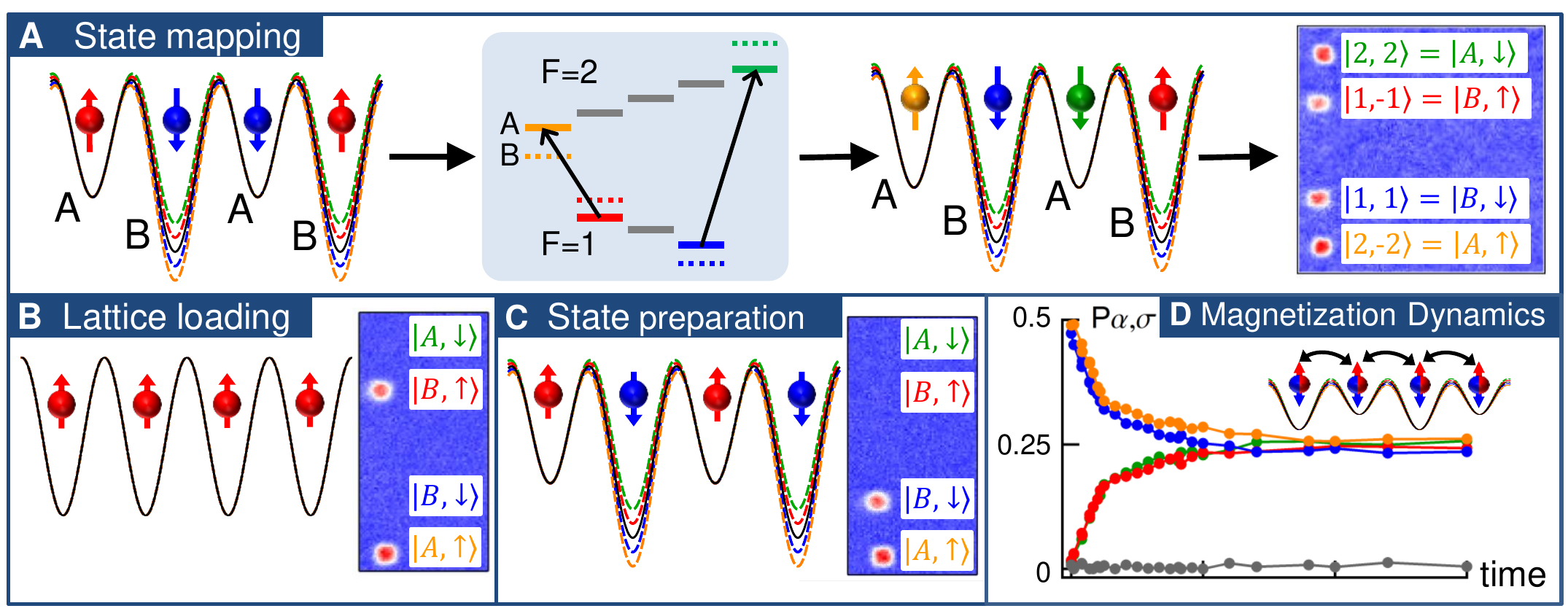}
\caption{ {\bf Schematic of experimental sequence.} {\bf (A)}  Spin/sub lattice mapping:  Atoms in $\left|\uparrow\rangle\right.$~(red) or $\left|\downarrow\rangle\right.$~(blue) occupy either the $A$ or $B$ sub-lattice (shown on the left). Applying a spin dependent addressing offset to the $B$ sub-lattice spectroscopically resolves the $A$ and $B$ sub-lattices (colored lines correspond to the potentials and energy levels seen by different hyperfine states). The $\left|A,\uparrow\rangle\right.$ and $\left|A,\downarrow\rangle\right.$ populations are microwave transferred to two different hyperfine states (yellow and green respectively), and the four mapped populations are measured by absorption imaging after Stern-Gerlach separation (shown on the right).  {\bf (B)} Initial lattice loading: A spin-polarized $\left|\uparrow\rangle\right.$~unit filled Mott insulator. {\bf (C)} Microwave state preparation: $B$ sites are microwave transferred from $\left|\uparrow\rangle\right.$ to $\left|\downarrow\rangle\right.$ using similar techniques to those employed for the state readout shown in {\bf (A)}. {\bf (D)} Time evolution: After the lattice is quenched to a specific configuration, the sub-lattice/spin populations are measured as a function of time (including the non-participating $m_F=0$ hyperfine state shown in gray).}
\label{fig:sequence}
\end{figure*}

Our experiment uses two hyperfine spin states~(denoted by $\uparrow$, $\downarrow$) of ultracold $^{87}$Rb atoms trapped in a dynamically controlled, 2D checkerboard optical lattice\cite{Sebby-StrableyPhysRevA2006}  comprised of two sub lattices $A$ and $B$ (Fig. 1a). For most experimental conditions presented here, our system is well described by a Bose-Hubbard model\cite{JakschPRL1998BHH} characterized by a nearest neighbor tunneling energy $J$, and an onsite interaction energy $U>0$. In addition, we use the lattice to apply an energy offset $\Delta_{\sigma} = \Delta + \delta_{\sigma}$ between the $A$ and $B$ sub-lattices, consisting of a spin-independent part $\Delta$ and a spin-dependent part $\delta_{\sigma}$ acting as a staggered magnetic field ($\sigma \in \{ \uparrow, \downarrow \}$)\cite{LeePRLSublatticeAddressingPRL2007}. All of these parameters can be dynamically controlled, which we exploit to prepare initial states with 2D anti-ferromagnetic order and to observe the resulting dynamics following a quench to different values of $J$, $U$, $\Delta$, and $\delta_{\sigma}$.

At unit filling, for $U\gg J$ and $\Delta_{\sigma} = 0$, double occupation at each site is allowed only virtually and the Bose-Hubbard model can be mapped onto a ferromagnetic Heisenberg model\cite{SpinModelDuanPRL2003,footy} with a nearest neighbor magnetic interaction strength $J_{\rm{ex}}$ that is second order in the tunneling energy. In the presence of hole impurities, first order tunneling (with the much larger energy scale $J$) must be included, which significantly modifies the dynamics even at low hole concentrations\cite{HildPRL2014}. The offset $\Delta_{\sigma}$ provides the flexibility to tune the relative importance of first order tunneling and second order superexchange processes. For example, below unit filling, if $|U-\Delta|\gg J$~and~$\delta_{\sigma}$, the Bose-Hubbard model can be mapped onto a bosonic $t$-$J$ model (a $J$-$J_{\rm{ex}}$ model in our notation, since $t$ represents time) with a staggered energy offset:
\begin{eqnarray}
H & = &  -J \sum_{\langle i,j \rangle, \sigma} a_{i\sigma}^\dagger a_{j\sigma} - \sum_{j \in A, \sigma} \Delta_{\sigma}
						a_{j\sigma}^\dagger a_{j\sigma}  \label{eq:tJmodel} \\
&  & - J_{\rm{ex}}\sum_{\langle i,j\rangle}\bm{S}_{i}\cdot\bm{S}_j
  - \sum_{\langle i,j,k\rangle,\sigma \sigma^\prime} V_{j} \left(a^{\dagger}_{i\sigma}\bm{\tau}^{}_{\sigma\sigma'}a^{\phantom\dagger}_{k \sigma^\prime}\right)\cdot\bm{S}_{j}. \nonumber
\end{eqnarray}
The local spin operators are defined as $\bm{S}_{i}=\frac{1}{2} \sum_{\sigma \sigma^\prime} a^{\dagger}_{i \sigma}\bm{\tau}_{\sigma \sigma^\prime}a^{\phantom\dagger}_{i \sigma^\prime}$, where $a^{\phantom\dagger}_{i \sigma}$ ($a^{\dagger}_{i \sigma}$) annihilates (creates) a hardcore boson of spin $\sigma$ on site $i$, and $\bm{ \tau}$ is a vector of Pauli matrices. The notation $\langle i, j \rangle$  indicates the sum over $i$ and $j$ is restricted to nearest neighbors, and $\langle i, j, k \rangle$  indicates the sum is restricted to sites $i, j, k$ such that $i \neq k$ are both nearest neighbors of $j$. The superexchange energy $J_{\rm{\small {ex}}} = 4 J^2 U /(U^2-\Delta^2)$ (Fig.~1b) can be either ferromagnetic ($U>\Delta$) or anti-ferromagnetic ($U<\Delta$)\cite{TrotzkyScience2008}. The last term describes hole-mediated exchange between sub lattices where an atom on site $k$ interacts via superexchange with an atom on site $j$, while simultaneously hopping to site $i$ (Fig.~1b). Here $V_j = V_{\pm} \equiv J^2/(U\pm \Delta )$, where $-$($+$) applies when $j \in A(B)$. 
In writing Eqn.~\ref{eq:tJmodel} we have ignored second-order processes\cite{FollingNature2007,LRTunnelMeinertSci2014} that conserve sub-lattice magnetization~\cite{Supplement}.  When $\left| \Delta \right| \lesssim J$, first order tunneling is resonant and dominates over hole mediated exchange.  For $\left| \Delta \right| \gg J$, however, first order tunneling is effectively suppressed, in which case the frequently ignored $V_j$ term plays an important role in hole motion and must be included. Similarly, superexchange is resonant when $\left|\delta_\sigma\right| \lesssim J_{\rm{ex}}$, but is suppressed when $\left|\delta_\sigma\right| \gg J_{\rm{ex}}$.  The values of $J$, $U$, $\Delta,$ and $\delta_{\sigma}$ are determined from an experimentally calibrated model of the lattice~\cite{Supplement}.  Inhomogeneity in the system, arising e.g. from trap curvature,  primarily enters via inhomogeneities in the parameters $\Delta$ and $\delta_\sigma$.

The experiments begin with $\lesssim 12 \times 10^3$ $^{87}$Rb atoms loaded into a square 3D optical lattice with one atom per site~\cite{Supplement}, initially spin polarized in the state $\left| \uparrow \rangle \right.$. We use the hyperfine states $ \left| \uparrow \rangle \right. \equiv \left| F\! =\! 1,\!m_F\! =\! -1 \rangle \right. $ and $ \left| \downarrow \rangle \right. \equiv \left| 1, +1 \rangle \right. $ to represent the pseudo-spin-1/2 system.
The 3D lattice is comprised of a vertical lattice along $z$ which confines the atoms to an array of independent 2D planes, along with the dynamic 2D checkerboard lattice in the $x$-$y$ plane.  The vertical lattice depth is typically $V_z = 35~E_\mathrm{R}$, held constant throughout the experiment, and the 2D lattice depth is initially $V_{xy} = 30~E_\mathrm{R}$ with no staggered offset, $\Delta_\sigma=0$~(The recoil energy $E_\mathrm{R}=h^2/(2m\lambda^2)$, $E_\mathrm{R}/h=3.47$~kHz, where $m$ is the mass of $^{87}$Rb and $\lambda= 813$~nm).  The atoms occupy roughly 13-15 2D planes, with the central plane containing 800-1100 atoms. 
The ratio of surface lattice sites to total lattice sites of the trapped cloud is $\approx$15~\% and sets a zero temperature lower bound for the number of sites with neighboring holes.
Based on spectroscopic measurements and assuming a thermal distribution~\cite{Supplement}, we estimate the hole density averaged over the entire cloud to be about 25~\%, and the hole density at the center of the cloud to be about 7~\%.

To measure the spin population independently on each sub-lattice, we map the four spin-spatial states $\left| A\uparrow \rangle \right. $, $\left| A\downarrow \rangle \right. $, $\left| B\uparrow \rangle \right. $ and $\left| B\downarrow \rangle \right. $ on to four distinct Zeeman states (Fig.~2a): By applying a large state-dependent offset $\delta_\sigma$ to all $\textit{B}$ sites we transfer the spectroscopically resolved $\textit{A}$-site atoms to two additional readout hyperfine states, $\left| A\uparrow \rangle \right. \rightarrow \left| 2, -2 \rangle \right. $ and $\left| A\downarrow \rangle \right.  \rightarrow \left| 2, +2 \rangle \right.  $\cite{LeePRLSublatticeAddressingPRL2007}. The four normalized populations $P_{\alpha,\sigma}$ ($\alpha \in \{A$, $B\}$) are measured with absorption imaging after Stern-Gerlach separation in a magnetic field gradient.

\begin{figure}
\centering\includegraphics [width=3.4in,angle=0] {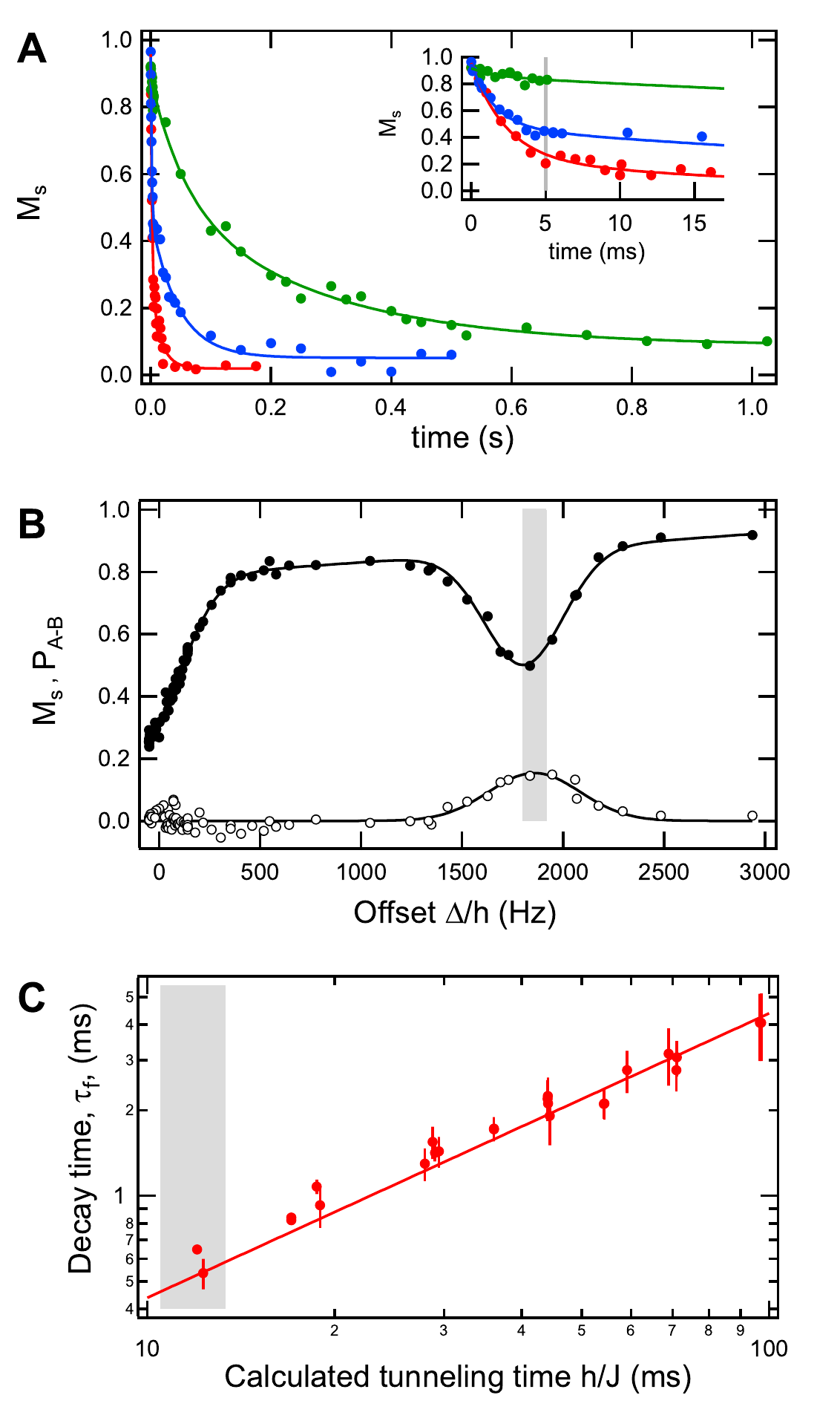}
\caption{{\bf Identification and control of tunneling} {\bf (A)} Decay of magnetization at a lattice depth of $15~E_{\mathrm{R}}$ for different offsets $\Delta/h$ of 1000~Hz (green), 300~Hz (blue), and -50~Hz (red). The inset shows the short time evolution, with two timescales ($\tau_f \approx$ 2~ms and $\tau_s\approx$ 50~ms),  both visible in the $\Delta/h$=~300~Hz~(blue) trace. The solid lines are double exponential fits. The vertical gray line indicates the fixed decay time at which the data in b) was taken.  {\bf (B)} Magnetization $M_s$ (filled circles) and sub lattice population $P_{A-B}$ (open circles) as a function of $\Delta$ at a fixed wait time of $5\:\mathrm{ms}\gtrsim \tau_f$ after the quench. The fast magnetization decay is resonant at $\Delta = 0$ and $\Delta = U$, while sub-lattice transport occurs near $\Delta=U$. The vertical gray band represents the calculated $U$ with an uncertainty due to parameter extraction from the two band model~\cite{Supplement}. {\bf (C)} The fast time scale, $\tau_f$, vs. calculated tunneling time scale $h/J$ for different lattice depths and $\Delta \!\lesssim \!J$. The solid line is $\tau_f = (h/J)/20$, and the gray band represents the uncertainty in the location of the 2D superfluid-insulator transition reported in Ref.\cite{SpielmanPRL2008}. (error bars represent the 1 standard deviation statistical uncertainties from fitting).}
\label{fig:InitialMeasurements}
\end{figure}

To perform the experiment, we start with a spin polarized configuration (Fig.~2b), and construct an initial state with staggered magnetization by applying the addressing offset $\delta_\sigma$ and transferring the $\textit{B}$-site atoms to $\left|\downarrow \rangle \right. $ ~(Fig.~2c). After returning $\delta_\sigma$ to zero we initiate dynamics by quenching to a given configuration with lattice depth $V_{xy}$ and offsets $\Delta$ and $\delta_\sigma$ (Fig.~2d). The ramp time for the quench of 200~$\mu$s was chosen to be fast with respect to subsequent dynamics but slow enough to avoid band excitation. After a variable hold time, we freeze the dynamics by raising $V_{xy}$ to $30~E_\mathrm{R}$ and read out the populations $P_{\alpha,\sigma}$, from which we determine the staggered magnetization $M_s$ and the sub lattice population difference $P_{A-B}$:
\begin{eqnarray}
M_s(t)& \equiv& P_{A,\uparrow}(t)+P_{B,\downarrow}(t)-P_{A,\downarrow}(t)-P_{B,\uparrow}(t), \label{eq:M_s} \nonumber \\
P_{A-B}(t)& \equiv &  P_{A,\uparrow}(t)+ P_{A,\downarrow}(t)- P_{B,\uparrow}(t) -P_{B,\downarrow}(t).
\label{eq:pa_b}
\end{eqnarray}
The exchange terms in Eqn.~\ref{eq:tJmodel} conserve $P_{A-B}$, while the first order tunneling does not, allowing for population transport between sub lattices. We also monitor the  total spin imbalance $P_{\uparrow -\downarrow}$ and the $m_F$=0 population to quantify unwanted spin-changing processes that drive the atoms out of the spin-$1/2$ manifold containing $\left|\uparrow\rangle \right.$ and $\left|\downarrow \rangle \right.$. We measure the time for depopulation of the spin-$1/2$ manifold to be greater than 6~s and the atom number lifetime in the lattice to be greater than 3~s.

\begin{figure}
\centering\includegraphics [width=3.4in,angle=0] {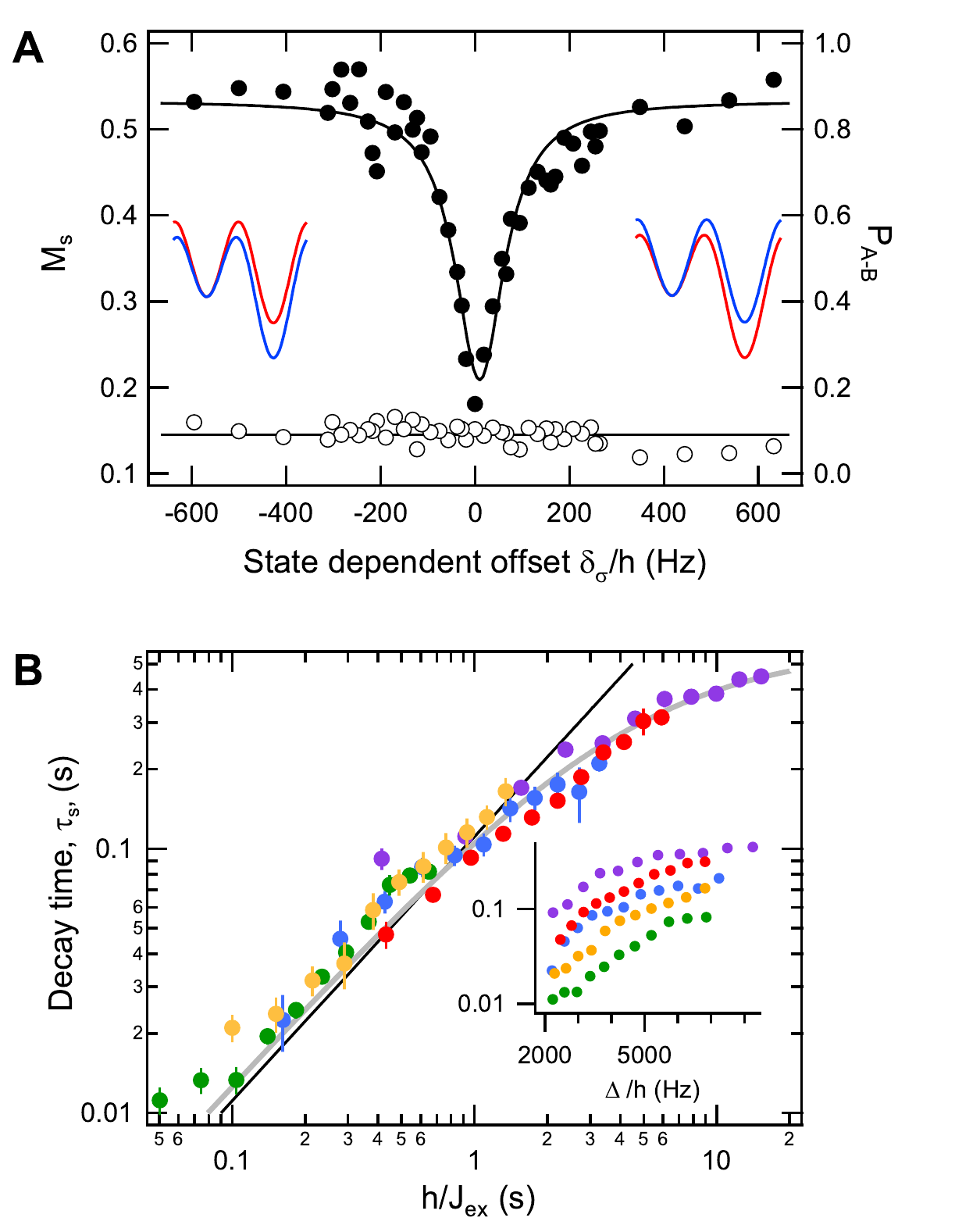}
\caption{{\bf Resonant superexchange} {\bf (A)} Magnetization~(filled circles) and sub lattice population difference~(open circles) at a fixed wait time of 70~ms vs. spin dependent energy offset $\delta_\sigma$ for $V_{xy}= 9.4~E_\mathrm{R}$ and $\Delta/h=4.3$~kHz. The initial and final states of the second order processes are resonant ($\delta_\sigma$=0), increasing the magnetization decay.  The inset lattice potentials show the sign change of $\delta_{\sigma}$ across resonance for a fixed offset $\Delta > U$.  {\bf (B)}  Measured slow decay time $\tau_s$ vs. calculated  superexchange time $h/J_{\rm{\small {ex}}}$:  The filled purple, red, blue, yellow, and green markers represent lattice depths of 14.7, 13.2, 11.3, 9.4, and 7.5~$E_\mathrm{R}$ respectively. Inset: measured slow decay rate vs applied staggered offset $\Delta$. The decay time scale, $\tau_s$, collapses with $h/J_{\rm{\small {ex}}}$ over roughly two orders of magnitude in $J_{\rm{\small {ex}}}$. The black line is a perturbative estimate of the scaling, which was checked in small systems by comparing to exact diagonalization averaged over hole-induced disorder~\cite{Supplement}. The gray line is a fit to a saturated linear dependence of $\tau_s$ on $h/J_{\rm ex}$ (see text). }
\label{fig:JexRes}
\end{figure}

For the lattice parameters studied in this paper, the magnetization dynamics is well described by exponential decay, with decay time scales ranging between 0.5~ms and 500~ms. Example decay curves are shown in Fig.~3a for a lattice depth $V_{xy}=15~E_\mathrm{R}$ and different offsets $\Delta$. For some $V_{xy}$ and $\Delta$  the exponential decay clearly occurs on two well separated time scales (Fig.~3a inset): a fast time scale, $\tau_f$, which dominates the behavior in shallow lattices when $\Delta \approx 0 $ or $\Delta \approx U$, and a slow time scale, $\tau_s$, which dominates the behavior in deep lattices with larger offset, $|\Delta|\gg U, J$. 

To investigate the faster time scale, we measure $M_s$ and $P_{A-B}$ at a fixed decay time for different $\Delta$, as shown in Fig.~3b for $V_{xy}=$15~$E_\mathrm{R}$.  The 5~ms decay time (indicated by the vertical line in  Fig.~3a) was chosen so that nearly all of the fast decay but little of the slow decay occurred. The fast magnetization decay reveals resonant features at $\Delta = 0$ and $U$, where the decay rate at $\Delta=0$ is twice as fast as at $\Delta=U$. In addition, $P_{A-B}$ shows sub-lattice transport from $B$ to $A$ sites at $\Delta =U$, indicative of resonant first order tunneling.  At $\Delta\approx0$ the demagnetizing sub lattice transport $B\rightarrow A$ and $A \rightarrow B$ are balanced. We theoretically estimate the expected width of the $\Delta=U$ resonance in $P_{A-B}$ to be $5 J/h = 110$~ Hz~\cite{Supplement}, which is significantly narrower than the 500~Hz width that we observe experimentally, suggesting inhomogeneous broadening.  
We note however, that the observed broadening is beyond what is expected from the measured trap curvature, and is inconsistent with estimates of light shift inhomogeneity from spectroscopic measurements~\cite{Supplement}. A residual $\delta_\sigma$ could account for the width. 

The measured decay times $\tau_f$ for a range of lattice depths are plotted against the calculated tunneling time $h/J$ in Fig.~3a, showing a decay rate linear in $J/h$, with $h/(J\tau_f)=20(2)$. This slope is comparable to a simple theoretical estimate taking only resonant tunneling into account, which predicts $h/(J\tau_f)\approx 2\pi\sqrt{2z}\approx18$, with $z=4$ the lattice coordination number.  Given the relatively large average hole density near the surface of the cloud, the agreement with a non-interacting estimate is not surprising, though we would expect interactions to reduce the decay rate.

To investigate the slow dynamics, we measure the magnetization decay time $\tau_s$ for $\Delta >U$, where first order tunneling was negligible and superexchange should dominate the dynamics.  To determine the dependence of $\tau_s$ on the spin-dependent staggered offset $\delta_{\sigma}$, we measure the remaining staggered magnetization $M_s$ and population difference $P_{A-B}$ after a fixed wait time $\approx \tau_s$, as shown in Fig.~4a for a lattice depth~9.4~$E_\mathrm{R}$ and offset $\Delta/h$=4.3~kHz. As expected for superexchange dynamics, the magnetization decay is resonant in $\delta_{\sigma}$.  The full width at half maximum of the Lorentzian fit to the resonance is 126(14)~Hz, considerably narrower than the observed tunneling resonances shown in Fig.~3b, and is most likely dominated by inhomogeneous broadening (second order exchange processes are sensitive to inhomogeneity in both $\Delta$ and $\delta_{\sigma}$). Figure~4a also shows that there is negligible sub-lattice transport associated with the demagnetization resonance. We note that at these values of $\Delta$, the ground state of the system would have all atoms on the lower sub-lattice, and the conservation of $P_{A-B}$ indicates that the spin dynamics occurs within a meta-stable manifold with respect to population.

Figure~4b shows the measured resonant decay times $\tau_s$ vs. calculated $h/J_{\rm{ex}}$ for different $V_{xy}$ and $\Delta$, with $\delta_\sigma=0$ and $\Delta$ chosen to be larger than $U$ but considerably less than the next excited band. The decay time $\tau_s$ scales with $h/J_{\rm{\small {ex}}}$ over two orders of magnitude in $J_{\rm{ex}}$. 
The solid gray line through the data is a fit to the expected linear dependence,  including a constant rate $\Gamma_{0}$ needed to capture the apparent saturation of $\tau_s$ at large~$h/J_{\rm{ex}}$: $\tau_s = ( A J_{\rm{ex}}/h + \Gamma_{0})^{-1}$, with $A = 7.8(4)$ and $\Gamma_{0}^{-1} =0.57(2)$~s.
A quantitative calculation of the decay rate in 2D, including the effects of holes, is extremely challenging.  However, the existence of a single energy scale contributing to the demagnetization in this regime justifies (at a qualitative level) a short-time perturbative treatment, from which we extract $A\approx 2\pi \sqrt{z/2} \approx 9$, in agreement with the experimentally measured value~\cite{Supplement}.
Surprisingly, this result is largely independent of the hole density, which can be attributed to the approximate cancellation of two competing effects of holes: they decrease the rate of superexchange dynamics, but simultaneously open new demagnetization channels through the final term in Eqn. (1).
The empirically determined time scale $\Gamma_{0}^{-1}$ is shorter than the spin depumping and number lifetimes, and may be related to the non-zero relaxation processes observed outside the $\delta_\sigma=0$ resonance in Fig.~4a. The mechanism for this off-resonant decay is not clear, but since the initial and final states differ in energy by significantly more than $J_{\rm{\small {ex}}}$ it must arise from energy-nonconserving processes such as noise assisted relaxation or doublon production\cite{StrohmaierPRL2010ElasticDoublonDecay}.
Corrections to $J_{\rm{\small {ex}}}$ due to excited band virtual processes\cite{Magnon_Boundstates_Fukuhara_Nat_2013Nat}, which we estimate to be of order 10-20~$\%$ at the largest $\Delta$ and smallest $V_{xy}$ shown in Fig. 4(b), may partially explain the observed saturation.

The scaling and resonant behavior of the fast and slow relaxation processes clearly reveal their origin as first-order tunneling and superexchange, respectively.  For $\Delta\gg J$, our experiment realizes an unusual situation in which tunneling is only active within a given sub-lattice and is comparable in strength to the superexchange coupling.  Both of these features---which are crucial to our ability to observe superexchange dominated dynamics---may have interesting implications for the relaxation of a doped antiferromagnetic state, potentially enabling a pre-thermalization scenario in which the spins can equilibrate in approximate isolation from the (typically higher entropy) motional degrees of freedom.  For smaller but non-zero $\Delta$, the ability to observe both tunneling and superexchange, often simultaneously and at experimentally accessible entropies, opens exciting opportunities to explore the non-equilibrium competition of spin and motion.  Understanding the detailed dynamics of this strongly-correlated, 2D quantum system is a formidable  challenge, which may require the development of new theoretical techniques.

This work was partially supported by the ARO's atomtronics
MURI, and NIST. M.F.F. and E.A.G. acknowledge support from the National Research Council Research Associateship program.  We thank; B. Grinkemeyer for his contributions to the data taking effort, E. Tiesinga and S. Paul for discussions about tight-binding models, and A. V. Gorshkov and S. Sugawa for a critical reading of the manuscript.

\newpage
\section*{\large Supplementary material}

\section*{Experimental Sequence and Methods}
 
All experiments begin with a $^{87}$Rb BEC with no discernible thermal fraction in the $|F=1,m_F=-1\rangle$ internal Zeeman state, optically trapped with trap frequencies $ \{ \nu_x, \nu_y, \nu_z \} = \{ 12,40,100\}$~Hz.  Control of the atom number, independent of trap parameters, is achieved by microwave removal of a fraction of atoms before the final stage of cooling. 
The BEC is then adiabatically loaded into a deep ($\approx$ 30~E$_r$) 3D $\lambda /2$-spaced lattice with  $\lambda=$813~nm, by loading the vertical lattice in $200$~ms and the 2D lattice in $100$~ms, starting 100~ms after beginning the vertical lattice ramp. The absence of doubly occupied sites is verified by number resolved microwave spectroscopy~\cite{GKCampbellScience2006} of the magnetically insensitive $|1,-1 \rangle \rightarrow |2,1 \rangle $ clock-transition near 0.323~mT using 80~ms pulses.  

Spectroscopic estimates of the total average trap inhomogeneity were made by measuring the  broadening of the clock transition and the state-dependent addressing transition ($|1,-1 \rangle \rightarrow |2,-2 \rangle $). These measurements indicate scalar and vector light shift inhomogeneity over the $\approx 10~\mu$m atom cloud is less than a percent of the total shift, about 800~Hz and 250~Hz respectively. Since $\Delta$ and $\delta_\sigma$ are only sensitive to light shift differences on the small length scale of $\lambda/2$, we expect the inhomogeneity in $A$-$B$ offsets to be substantially smaller than the measured globally averaged inhomogeneity. 

We measure the average number of holes throughout the atom cloud by performing number resolved microwave spectroscopy~\cite{GKCampbellScience2006} after merging neighboring pairs of sites into one site. Any pair of sites that contains a hole is counted as having one atom, and comparing the merged two-atom signal to the merged one-atom signal gives a measure of the average hole density. We estimate the central hole density by assuming a Fermi-Dirac distribution in the harmonic trap (under the assumption that there are no doubly occupied sites and the system is in thermal equilibrium) having a chemical potential and temperature that matches the measured number and average merged one atom fraction. 

\section*{Tight Binding Parameters}
The tight binding parameters $U$, $J$, $\Delta$ and $\delta_\sigma$ are determined from an experimentally calibrated model of the 2D lattice potential. The details of the checkerboard optical lattice are described in Ref.~\cite{Sebby-StrableyPhysRevA2006}, and we only give a brief description here relevant for extracting tight binding parameters. The lattice is generated from a single laser beam folded to produce four interfering beams propagating along the $x$ and $y$ directions, resulting in a position-dependent total field
\begin{eqnarray}
\vec{E}_{\rm latt}(x,y) &= (E_1 \hat{e}_1 e^{-i k x} + E_2 \hat{e}_2 e^{-i k y} +\\ \nonumber & E_3 \hat{e}_3 e^{ i k y} +E_4 \hat{e}_4 e^{ i k x} ),\label{eqn:Efield}
\end{eqnarray}
where $k= 2 \pi/\lambda$, $\lambda=813$~nm is the wavelength of the lattice light, and $E_{i}$ are the single beam field amplitudes. The orientation and phase of the complex unit vectors $\hat{e}_i$ are controlled with electro-optic modulators (EOMs). The local intensity $I_{\rm latt} = c \epsilon_0 |\vec{E}_{\rm latt}|^2 $ and circular polarization $i ( \vec{E}^*_{\rm latt} \times \vec{E}_{\rm latt})$ give rise to a scalar light shift potential $V_{\rm latt}(x,y)$ and effective Zeeman field $\vec{B}_{\rm eff}(x,y)$ respectively~\cite{deutsch98a}. The lattice can be tuned between a square lattice with $\lambda/2$ periodicity along $x$ and $y$, and a square lattice with $\lambda/\sqrt{2}$ periodicity along $x+y$ and $x-y$. The spin-dependent lattice potential is calibrated for our geometry using the measured transmission losses, the calibrated polarization responses of the EOMs (including hysteresis), the measured deviation from orthogonality of the beams along $x$ and $y$, microwave spectroscopy~\cite{LeePRLSublatticeAddressingPRL2007}, diffraction phase winding measurements~\cite{SebbyStrableyPhysRevLett2007,Sebby-StrableyPhysRevA2006} and pulsed diffraction to calibrate the depth~\cite{Gupta2001}. The input field is calibrated in terms of the measured lattice depth $E_1=(1/2) \sqrt{(V_{xy}/E_\mathrm{R})}$, where the lattice depth in recoil units $V_{xy}/ E_\mathrm{R}$ is determined for the configuration of a square $\lambda/2$ lattice.

\begin{figure}[t!]
\includegraphics[width=1.0\columnwidth]{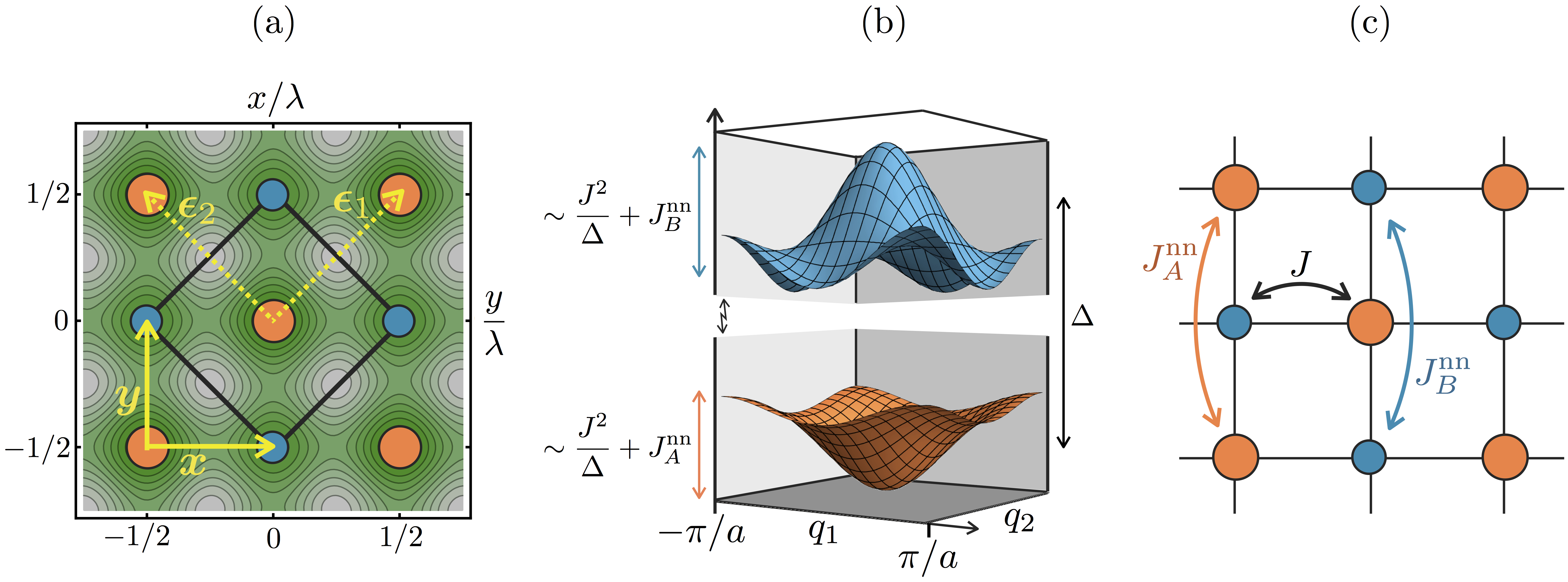}
\caption{(a) The 2D optical lattice potential used for dynamics experiments, showing a unit cell outlined in black.  (b) Typical example of the lowest two bands of the lattice (in the limit $\Delta\gg J$) in the first Brillouin zone, which is reciprocal to the unit cell drawn in (a).  (c) Lattice sites divided into two sub lattices, and examples of the matrix elements entering the tight-binding model used to calculate the band structure in \eref{SuppEq:tb}.}
\label{SuppFig:lattice}
\end{figure}

The full lattice potential, including imperfections, is used in the calculation of Bose-Hubbard parameters described below, but the approach we take is simplest to describe without  lattice imperfections. In the absence of transmission losses or birefringence ($E_i = E_{xy}$), the scalar part can be written as
\begin{align}
V_{\rm latt}(x,y) &= V_\parallel(\theta_1) \left(\cos{2 k x}+\cos{2 k y} \right)+\nonumber \\ 
&V_\perp(\theta_1) \left[ \cos{(k x -\theta_2)} + \cos{k y} \right]^2. 
\end{align}
Here $\theta_1$ and $\theta_2$ are controlled by two separate EOM's, and $V_\parallel(\theta_1) = V_{xy}(1/2)\cos^2{\theta_1}$ and $V_\perp(\theta_1) = V_{xy} \sin^2{\theta_1}$ are parameterized by $V_{xy}$ determined when $\theta_1=0$. (For $\theta_1= \pi/2$, the total lattice depth would be $4 V_{xy}$.) In the limit of small $\theta_1$, $V_\perp \ll V_\parallel$. If in addition $\theta_2=0$ or $\pi$, the lattice can be described as a square lattice of spacing $\lambda/2$ with a staggered offset $\Delta\approx 4 V_{xy} \sin^2{\theta_1}$~(Fig.\,\ref{SuppFig:lattice}a).  We use experimentally measured values of $\Delta$ under different conditions to calibrate the lattice model and the dependence of $\Delta$ on $\{V_{xy},\theta_1,\theta_2\}$. 
The effective field $\vec{B}_{\rm eff}(x,y)$ lies in the $xy$ plane and is similarly controlled by $\theta_1$ and $\theta_2$. In the presence of a large bias field $\vec{B}_0\gg \vec{B}_{\rm eft}$, the total spin-dependent staggered offset $\delta_\sigma\propto | \vec{B}_0+  \vec{B}_{\rm eff}|$ depends on the relative angle between $\vec{B}_0$ and $\vec{B}_{\rm eff}$:
\begin{equation}
\left| \vec{B}_0+  \vec{B}_{\rm eff}(x,y) \right| \approx \left| \vec{B}_0\right| 
+ \vec{B}_{\rm eff}(x,y) \cdot \left( \frac{\vec{B}_0}{\left| \vec{B}_0 \right| } \right).
\end{equation}
We control the size of the spin-dependent shift $\delta_\sigma$ by changing the orientation of $\vec{B}_0$ with respect to the lattice, so that $\delta_\sigma \approx 0$ when $\vec{B}_0 \perp \vec{B}_{\rm eff}$. Microwave spectroscopy is used to calibrate $\delta_\sigma$ as a function of $\{V_{xy},\theta_1,\theta_2,\vec{B}_0\}$.

The tunneling parameter $J$ is extracted from band structure calculations based on the full lattice potential.   Except in the limit when $\theta_1=0$, the potential is not separable in Cartesian coordinates and the band structure calculations must be carried out in 2D.  The primitive unit cell of our lattice, shown in Fig.~\,\ref{SuppFig:lattice}a, is spanned by primitive vectors $\bm{\epsilon}_1$ and $\bm{\epsilon}_2$ of length $a=\lambda/\sqrt{2}$.  A typical example of the lowest two bands, $E_-(\bm{q})$ and $E_+(\bm{q})$ (where $\bm{q}=\{q_1,q_2\}$ has components conjugate to $\bm{\epsilon}_1$ and $\bm{\epsilon}_2$), is shown in Fig.\,\ref{SuppFig:lattice}b.  In order to determine the inter-sublattice tunneling matrix elements, we first calculate the lowest two bands of a suitable two-band tight-binding model analytically.  

In the limit $\Delta\gg J$, the bandwidth contribution to either of the lowest two bands from the direct hopping $J$ can be estimated perturbatively as $\sim J^2/\Delta$, which scales with $V_{xy}$ similarly  to the next-nearest-neighbor tunneling amplitudes directly connecting sites of the $A$($B$) sublattice, denoted $J_{A(B)}^{\rm nn}$ (Fig.\,\ref{SuppFig:lattice}c). As a result, an accurate tight binding model must include $J_A^{\rm nn}$ and $J_B^{\rm nn}$, in which case we find tight binding bands
\begin{align}
&\mathcal{E}_{\pm}(\bm{q}:\Delta,J,J_{A}^{\rm nn},J_{B}^{\rm nn}) = \Delta/2 \nonumber \\
&-2(J_{A}^{\rm nn}+J_{B}^{\rm nn})\cos q_1a\cos q_2a\nonumber \\
&\pm(\big(\Delta/2-2(J_{A}^{\rm nn}-J_{B}^{\rm nn})\cos q_1a\cos q_2a\big)^2 \nonumber \\
& + 4J^2\big(1+\cos q_1a+\cos q_2a+\cos q_1a\cos q_2a\big))^{1/2} .
\label{SuppEq:tb}
\end{align}

We extract the dependence of $J$ on $\{V_{xy},\theta_1,\theta_2\}$ by fitting $\mathcal{E}_{\pm}(\bm{q}:\Delta,J,J_{A}^{\rm nn},J_{B}^{\rm nn})$ to the numerically calculated $E_{\pm}(\bm{q},V_{xy},\theta_1,\theta_2)$.
With the next-nearest-neighbor tunnelings included, the fits typically produce a Brillouin zone averaged fractional error in the band energies on the order of $10^{-3}$. Under almost all conditions in the paper, the extracted $J$ is essentially independent of $\Delta$ and can be determined from the $\Delta=0$ lattice with equivalent depth $V_{xy}$.  The interaction energy $U$ is given by $U = g \int{d^3r \left| \phi(\bm{r})\right|^4  }$ where $g = 4 \pi \hbar^2 a_s /m$, $a_s$ is the $s$-wave scattering length, $m$ is mass of $^{87}$Rb, and $\phi(\bm{r})$ is the localized Wannier function on a lattice site. Assuming $\phi$ is a gaussian wave function (appropriate for a harmonic expansion of the lattice site), $U = g/((2 \pi)^{3/2} a_x a_y a_z)$ where $a_{x,y,z}$ are the harmonic oscillator lengths associated with the local lattice site curvature. The anharmonicity on a lattice site in a square $\Delta=0$ lattice can be approximately accounted for by using a Gaussian wave function with a modified width of $a_x = (\lambda/2\pi )/\sqrt{\sqrt{V_{xy}} -1/2}$.  A calculation of $J_{\rm ex}$ that takes into account the $\Delta \neq 0 $ impact on $U$ deviates by less than 6~\% from the $\Delta=0$ value over the range of $\Delta$ considered here, and we use the simple $\Delta=0$ analytical expression for $U$ described above.

\section*{Superexchange Timescale Estimates}
When $|U-\Delta |\gg J$, (and assuming that $\delta_{\sigma}\ll\Delta$), double occupancies are forbidden and the Bose-Hubbard model can be mapped onto a bosonic $t$-$J$ model with a sub-lattice detuning,
\begin{align}
\label{eq:sup_ex_1}
H&=-J\sum_{\langle i,j\rangle,\sigma}a^{\dagger}_{i\sigma}a^{}_{j\sigma}-\sum_{j\in A,\sigma}\Delta_{\sigma}a^{\dagger}_{j\sigma}a^{}_{j\sigma}\nonumber\\
&-J_{\rm ex}\sum_{\langle i,j\rangle}\bm{S}_{i}\cdot\bm{S}_j 
-\sum_{\langle i,j,k\rangle,\sigma\sigma'}V_j \left(a^{\dagger}_{i\sigma}\bm{\tau}_{\sigma\sigma'}a^{\phantom\dagger}_{k\sigma'}\right)\cdot\bm{S}_{j} \nonumber \\
& -\frac{3}{2}\sum_{\langle i,j,k\rangle,\sigma} V_j a^{\dagger}_{i\sigma}a^{\phantom\dagger}_{k\sigma}n_{j}.
\end{align}
Here
\begin{align}
J_{\rm ex}=\frac{4J^2 U}{U^2-\Delta^2}~~~~~~V_{j}=\frac{J^2}{U-\kappa_j\Delta},
\end{align}
and $\kappa_j=+1$ or $\kappa_j=-1$ depending on whether $j$ is contained in the $A$ or $B$ sub lattice, respectively.  For all of the data in Fig.\,4 of the manuscript, $\Delta\gg J$.  As a result, terms which change the sub lattice population (i.e. the remaining hopping processes which move a single atom between two adjacent and otherwise empty lattice sites) can also be integrated out at second order in the small parameter $J/\Delta$, yielding
\begin{align}
\label{SuppEq:sup_ex_2}
H&=-J_{\rm ex}\sum_{\langle i,j\rangle}\bm{S}_{i}\cdot\bm{S}_j-\!\!\!\sum_{\langle i,j,k\rangle,\sigma\sigma'}\tilde{V}_{j} \left(a^{\dagger}_{i\sigma}\bm{\tau}_{\sigma\sigma'}a^{\phantom\dagger}_{k\sigma'}\right)\cdot\bm{S}_{j} \nonumber \\
&-\frac{3}{2}\sum_{\langle i,j,k\rangle,\sigma} \tilde{V}_{j}a^{\dagger}_{i\sigma}a^{\phantom\dagger}_{k\sigma}n_{j}+\!\!\!\sum_{\langle i,j,k\rangle,\sigma}\frac{J^2}{\kappa_j\Delta}a^{\dagger}_{i\sigma}a^{\phantom\dagger}_{k\sigma}.
\end{align}
Here $\tilde{V}_{j}=J^2/(U-\kappa_j\Delta)+J^2/(\kappa_j\Delta)$ is modified from the previously defined $V_{j}$ to account for second order hole motion consistent with the hardcore constraint.  We note that the first two terms couple states with different sub lattice magnetization, while the second two do not.

\begin{figure}[t!]
\includegraphics[width=0.7\columnwidth]{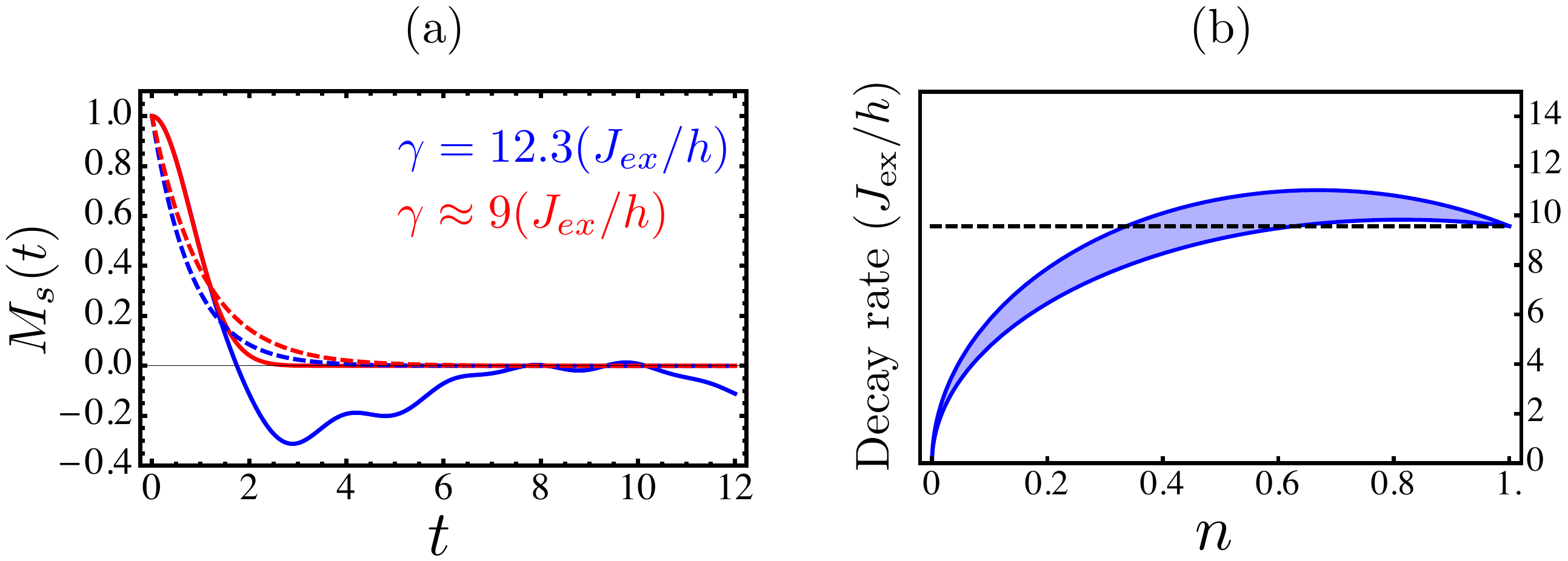}
\caption{(a) Demagnitezation dynamics from a unit-filled Ne\'{e}l state.  The blue solid line is from exact diagonalization of a $4\times 4$ plaquette with periodic boundary conditions, and the blue dotted line is a fit to an exponential $e^{-\gamma t}$.  The red solid line is from perturbation theory, and the red dotted line is once again a fit to an exponential~(this procedure gives the slope of the gray line plotted in Fig.~4b).  (b) Demagnetization rate calculated at second-order in short-time perturbation theory, as a function of density.  The blue shaded region reflects a range of the ratio $1<\Delta/U<5$, which encompasses all data points shown in Fig. 4 of the manuscript, while the black-dashed line shows the superexchange time-scale at unit filling.}
\label{SuppFig:plots}
\end{figure}

A priori, for a finite hole density (i.e. when the demagnetization channel associated with the second term in \eref{SuppEq:sup_ex_2} is active), the demagnetization rate at short times does not need to scale with $J_{\rm ex}$.  However, we find that over a broad range of densities, the short-time demagnetization rate does indeed scale with $J_{\rm ex}$ to a good approximation, as we now show.  Because the initial state is an eigenstate of the staggered magnetization operator
\begin{align}
\hat{M}_s=2\Big(\sum_{j\in A}S_j^z-\sum_{j\in B}S_j^z\Big)/N,
\end{align}
with $N$ the total number of atoms, the initial decay of the magnetization must be quadratic in time.  Defining $M_s(t)=\langle \psi(t)| \hat{M}_s|\psi(t)\rangle$, we can choose to expand the 
magnetization as $M_s(t)=\exp[-m_2 t^2+\mathcal{O}(t^3)]$, where
\begin{align}
m_2=\frac{1}{2}\langle[H,[H,\hat{M}_s]]\rangle/\hbar^2.
\end{align}
Working to second order, we then extract a time-scale by fitting $e^{-m_2 t^2}$ to an exponential $e^{-\gamma t}$, giving $\gamma\approx \sqrt{m_2}$.  This approximation is only expected to give a qualitative estimate of the decay time scale, valid under the assumption that a significant portion of the decay occurs at or below the timescale $\hbar/J_{\rm ex}$.  However, this estimate is fairly accurate when compared to exact diagonalization results for a superexchange model on a $4\times 4$ plaquette (see Fig.\,\ref{SuppFig:plots}a).

Under the assumption that holes are distributed randomly, extensive but straightforward algebra leads to
\begin{equation}
\gamma\approx\frac{1}{\hbar}\sqrt{\frac{n z}{2}\times J_{\rm ex}^2+2n(1-n) z(z-1) \times\Big((\tilde{V}_{+})^2+(\tilde{V}_{-})^2\Big)},
\end{equation}
where $n$ is the density, $z=4$ is the lattice coordination number, and $\tilde{V}_{+}(\tilde{V}_{-})$ is equal to $\tilde{V}_{j}$ when $j$ is contained in the $B(A)$ sub lattice.  This result is plotted (in units of $J_{\rm ex}/h$) for a range of values of $U/\Delta$ in Fig.\,\ref{SuppFig:plots}b, where we see that for a broad range of densities ($n\gtrsim 0.3$) the rate is in good quantitative agreement with the one extracted at unit filling $\big(\gamma\approx(J_{\rm ex}/\hbar)\sqrt{z/2}, ~\rm{dashed~line}\big)$, where $J_{\rm ex}$ is the only relevant energy scale in the Hamiltonian.

\section*{The $\Delta=U$ population imbalance resonance}

When $U\gg J$ and the staggered offset is near $\Delta=U$, all dynamics occurs within the subspace where the $A$ sub lattice has either one or two atoms on each site, while every site of the $B$ sub lattice has either one or zero atoms.  If, for simplicity, we ignore the spin degrees of freedom, and consider doubly occupied $A$ sites to be particles and singly occupied $A$ sites to be holes, the density degrees of freedom map onto hardcore spinless bosons hopping with strength $J$,
\begin{align}
H_{\rm res}=-J\sum_{\langle i,j\rangle}b^{\dagger}_{i}b^{\phantom\dagger}_{j}+\tilde{\Delta}\sum_{j\in A}b^{\dagger}_jb^{\phantom\dagger}_j.
\end{align}
Here the operator $b(b^{\dagger})$ annihilates(creates) a hardcore boson, the sub lattice detuning is related to the actually staggered offset by $\tilde{\Delta}=U-\Delta$, and the initial state contains a single boson on every site of the $B$ sub lattice and none on the $A$ sub lattice.  Note that the tunneling energy for the hardcore bosons in this model is ambiguous up to a factor of $\sqrt{2}$, since the matrix element for an atom to hop from the $B$ to the $A$ sub lattice depends on whether the ``hole'' on the $A$ sub lattice has the same spin as the hopping particle.  Assuming a translationally invariant lattice with $\mc{N}$ sites, the average hardcore boson density on the $A(B)$ sub lattice is given by $\tilde{n}_{A(B)}=(2/\mc{N})\sum_{j\in A(B)}\langle b^{\dagger}_j b^{\phantom\dagger}_j\rangle$.  The tilde in $\tilde{n}_{A(B)}$ indicates that these are not the densities of the physical atoms, which we denote by $n_A$ and $n_B$, and are related by $n_{A}=\tilde{n}_A+1$ and $n_B=\tilde{n}_B$.  At unit filling ($n_A+n_B=2$), the experimentally-measured population difference, $P_{A-B}(t)=\frac{1}{2}[n_A(t)-n_B(t)]$, can be expressed as $P_{A-B}(t)=1-n_B(t)=1-\tilde{n}_B(t)$.  As described above, the initial state in the hardcore boson picture has $\tilde{n}_B(0)=1$, and hence $P_{A-B}(0)=0$.  The steady-state (at $\tilde{\Delta}=0$, i.e. on resonance) must have $\tilde{n}_A=\tilde{n}_B=1/2$, and hence $P_{A-B}=1/2$.

The steady-state population imbalance is given by $P^{\rm ss}_{A-B}(\tilde{\Delta})=1-\tilde{n}_B$ for $t\gg \hbar/J$, and we would like to know how $P_{A-B}^{\rm ss}(\tilde{\Delta})$ depends on the sub lattice detuning $\tilde{\Delta}$ (this is what is measured experimentally (blue data points) in Fig. 3(b) of the manuscript).  A simple estimate can be obtained by just relaxing the hardcore constraint, in which case the problem becomes non-interacting.  We can then obtain $\tilde{n}_B(t)$ by solving the dynamics of a single atom starting on the $B$ sub lattice.  Working in quasi-momentum space, the single-particle eigenstates in a staggered lattice can be obtained by diagonalizing the matrix
\begin{equation}
\mathcal{H}(\bm{q})=\left(
\begin{array}{cc}
\tilde{\Delta}/2 &  \varepsilon(\bm{q})  \\
\varepsilon(\bm{q})  &   -\tilde{\Delta}/2  \\   
\end{array}
\right),
\end{equation}
where $\varepsilon(\bm{q})=2J(\cos q_x a+\cos q_y a)$ is the single-particle spectrum at zero detuning.  Our initial state is evenly distributed across the Brillouin zone, but at every $\bm{q}$ it will be decomposed differently onto the eigenvectors of the above matrix.  Obtaining $\tilde{n}_B(t)$ simply requires solving a standard off-resonant Rabi problem at each $\bm{q}$, with generalized Rabi frequency $\Omega(\bm{q})=\sqrt{\tilde{\Delta}^2+4\varepsilon(\bm{q})^2}$.  Integrating such solutions over the first Brillouin zone, we obtain
\begin{equation}
\tilde{n}_{B}(t)=\frac{1}{A_{\rm bz}}\int_{\rm bz}d^2q\frac{\Omega(\bm{q})^2+2\varepsilon(\bm{q})^2\left(\cos\Omega(\bm q) t-1\right)}{\Omega(\bm{q})^2},
\end{equation}
where $A_{\rm bz}$ is the first Brillouin zone area.  The size of the time-dependent term in this integral decreases at large $t$ as $1/\sqrt{t}$ (as can be seen from a stationary phase approximation), and so only the time-independent piece survives at long time, giving
\begin{equation}
P^{\rm ss}_{A-B}(\tilde{\Delta})=\frac{1}{A_{\rm bz}}\int_{\rm bz}d^2q\frac{2\varepsilon(\bm{q})^2}{\Omega(\bm{q})^2}.
\end{equation}
Taking the integral numerically, we find that the full-width at half-max of this resonance is approximately $5J$, which is significantly narrower than the experimentally measured feature.

\bibliography{Bigbib}

\begin{thebibliography}{35}
\bibitem[{foo()}]{footy}
\bibinfo{note}{Strictly speaking the validity of Eq.~1 also requires
  $|\Delta-U/2| \gg J^2/\Delta$ and $|\Delta-3U/2| \gg J^2/|U-\Delta|$, to
  avoid narrower second order resonances which are associated with double and
  triple occupancy, respectively.}
\end{thebibliography}

\end{document}